\documentclass[aps,prc,amsmath,amssymb,showpacs,superscriptaddress,twocolumn,footinbib,floatfix]{revtex4}
\usepackage{longtable,graphicx,dcolumn,epsfig,enumerate}
\begin{document}
\title{Dipole strength in $^{144}$Sm studied via ($\gamma$,n),
($\gamma$,p) and ($\gamma,\alpha$) reactions}
\author{C. Nair}
\altaffiliation{Present address: Nuclear Engineering Division, 
Argonne National Laboratory, Argonne, Illinois 60439, USA.}
\affiliation{Institut f\"ur Strahlenphysik, Forschungszentrum
             Dresden-Rossendorf, D-01314 Dresden, Germany}
\author{A. R. Junghans}
\affiliation{Institut f\"ur Strahlenphysik, Forschungszentrum
             Dresden-Rossendorf, D-01314 Dresden, Germany}
\author{M. Erhard}
\altaffiliation{Present address: Istituto Nazionale di Fisica
Nucleare, Sezione di Padova, 35131 Padova PD, Italy.}
\affiliation{Institut f\"ur Strahlenphysik, Forschungszentrum
Dresden-Rossendorf, D-01314 Dresden, Germany}
\author{D. Bemmerer}
\affiliation{Institut f\"ur Strahlenphysik, Forschungszentrum
             Dresden-Rossendorf, D-01314 Dresden, Germany}
\author{R. Beyer}
\affiliation{Institut f\"ur Strahlenphysik, Forschungszentrum
             Dresden-Rossendorf, D-01314 Dresden, Germany}
\author{E. Grosse}
\affiliation{Institut f\"ur Strahlenphysik, Forschungszentrum
             Dresden-Rossendorf, D-01314 Dresden, Germany}
\affiliation{Institut f\"ur Kern- und Teilchenphysik,
             Technische Universit\"at Dresden, D-01062 Dresden, Germany}
\author{K. Kosev}
\affiliation{Institut f\"ur Strahlenphysik, Forschungszentrum
             Dresden-Rossendorf, D-01314 Dresden, Germany}
\author{M. Marta}
\affiliation{Institut f\"ur Strahlenphysik, Forschungszentrum
             Dresden-Rossendorf, D-01314 Dresden, Germany}
\author{G. Rusev}
\altaffiliation{Present address: Department of Physics, Duke
University and Triangle Universities Nuclear Laboratory, Durham, NC
27708, USA.} \affiliation{Institut f\"ur Strahlenphysik,
Forschungszentrum Dresden-Rossendorf, D-01314 Dresden, Germany}

\author{K. D. Schilling}
\affiliation{Institut f\"ur Strahlenphysik, Forschungszentrum
             Dresden-Rossendorf, D-01314 Dresden, Germany}
\author{R. Schwengner}
\affiliation{Institut f\"ur Strahlenphysik, Forschungszentrum
             Dresden-Rossendorf, D-01314 Dresden, Germany}
\author{A. Wagner}
\affiliation{Institut f\"ur Strahlenphysik, Forschungszentrum
             Dresden-Rossendorf, D-01314 Dresden, Germany}

\date{\today}

\begin{abstract}
Photoactivation measurements on $^{144}$Sm have been performed with
bremsstrahlung endpoint energies from 10.0 to 15.5 MeV at the
bremsstrahlung facility of the superconducting electron accelerator
ELBE of Forschungszentrum Dresden-Rossendorf. The measured
activation yield for the $^{144}$Sm($\gamma$,n) reaction is compared
with the calculated yield using cross sections from previous
photoneutron experiments. The activation yields measured for all
disintegration channels $^{144}$Sm($\gamma,n$), ($\gamma,p$) and
($\gamma,\alpha$) are compared to the yield calculated by using
Hauser-Feshbach statistical models. A new parametrization of the
photon strength function is presented and the yield simulated by
using the modified photon strength parameters are compared to the
experimental data.
\end{abstract}
\maketitle

\section{Introduction}
\label{intro}

The nuclei heavier than iron $(Z>26)$ are synthesized mainly by
neutron-capture reactions- the astrophysical r- and s-processes.
There are about 35 neutron deficient stable isotopes between Se and
Hg that are shielded from the rapid neutron capture by stable
isobars. These nuclei, classically referred to as the p-nuclei, are
produced via chains of photodisintegrations such as ($\gamma,n$)
($\gamma,p$) and ($\gamma,\alpha$) on r- or s-seed
nuclei~\cite{B2FH,Lambert1992}. The best possible sites that have
been proposed for the production of p-nuclei are the O/Ne-rich
layers of type II supernova explosions~\cite{Arnould2003}.

For the calculation of p-process abundances, a comprehensive network
involving thousands of reaction rates is necessary. Despite the
efforts in recent years, the experimental information on the
reaction rates involved in the p-process nuclear flows is very
scarce. The p-process reaction rates presently used in the
astrophysical network calculations are based on the cross sections
obtained from Hauser-Feshbach statistical model calculations.

For the network calculations involving the p-nuclei, the precise
knowledge of the photon strength especially in the close-threshold
regions is very important. In contrast to the rather detailed
knowledge~\cite{Dietrich1989} of the photon strength in the
isovector giant dipole resonance (GDR) region well above the
particle separation energies, the corresponding information for the
excitation region near threshold is still surprisingly uncertain due
to the much smaller cross section. Accurate information on weaker
channels with protons or alpha particles in the exit channels is
also necessary.

The experimental information available for the p-nucleosynthesis so
far is either from the reactions that involve neutrons or charged
particles. The work described here presents experimental data for
the photodisintegration of one nucleus into four different channels
as observed in a wide range of bremsstrahlung endpoint energies, i.e, the
$^{144}$Sm($\gamma,n$) to $^{143}$Sm and $^{143m}$Sm, the
$^{144}$Sm($\gamma,p$) and $^{144}$Sm($\gamma,\alpha$) reactions.
The nuclide $^{144}$Sm has been discussed in the frame of the
p-process chronometer $^{146}$Sm~\cite{Audo72,Andr06}. There were
several efforts to determine the $^{146}$Sm/$^{144}$Sm production
ratio experimentally which varies due to uncertainties in different
inputs entering into the calculation, see Ref.~\cite{Somo98}. The
nuclear-physics-related uncertainties in the p-process model
predictions were investigated recently by Rapp $\textit{et
al.}$~\cite{Rapp06}. A list of the most critical p-process reaction
rates (with their respective inverse reactions), which will
influence the final p-abundances, have been given (see Table 2 and
Table 3, Ref. \cite{Rapp06}). In general, a systematic investigation
of the ($\gamma$,$\alpha$) reactions in the mass region $A$$\geq$140
is found to be very important. Therefore, experimental information
on the photodisintegration rates in $^{144}$Sm helps to reduce the
inherent nuclear model uncertainties.

The photoactivation experiments presented here cover a substantial
range of excitation. The experimental setup is described in
Sec.~\ref{sect:setup} and Sec.~\ref{sect:datanal} explains the
method of photoactivation yield determination. The experimental
results of the $^{144}$Sm($\gamma,n$), $^{144}$Sm($\gamma,p$) and
$^{144}$Sm($\gamma,\alpha$) reactions are discussed in detail under
Sec.~\ref{sect:sm-yield-results}. The photoactivation experiments
are performed such that an absolute yield can be extracted which
allows the derivation of an absolute strength. This strength
function is used in statistical calculations of Hauser-Feshbach type
as an input parameter. This point is worked out in detail in
Sec.~\ref{sect:sm-gnyield-modelcalc} of this paper.

\section{Experimental Setup}\label{sect:setup}

The photodisintegration experiments discussed here were performed at
the bremsstrahlung facility of the superconducting electron
accelerator ELBE (Electron Linear accelerator of high Brilliance and
low Emittance) of Forschungszentrum Dresden-Rossendorf. At ELBE,
bremsstrahlung endpoint energies up to 20 MeV and average currents
up to 1 mA are available which is appropriate for probing
photon-induced reactions. The bremsstrahlung facility has been
extensively used for photon scattering as well as photoactivation
studies~\cite{Rusev2006,Schwengner2007,Schwengner2008,Rusev2008,Rusev2009,Benouaret2009,Nair2008}.
The experimental setup at ELBE has been described in detail
elsewhere~\cite{Schwengner2005,Wagner2005}.

\begin{figure}
\begin{center}
\includegraphics[width=8cm,angle=0]{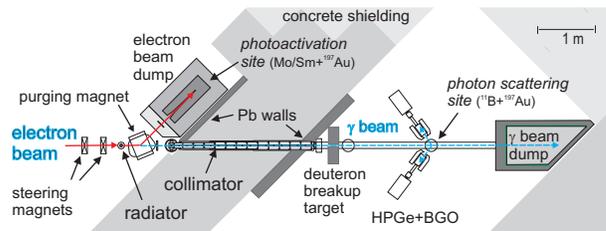}
%original:C:\Documents and Settings\chithra.FZR\Desktop\publications\POS-NIC10\usedfiles\fig1\brems-facility.eps
\caption{\label{fig:experimentalsetup}(Color online) Photoactivation
setup at ELBE accelerator~\cite{Schwengner2005}. The electron beam
is deflected from the main beam line and creates bremsstrahlung in
the radiator. There are two target sites: at the photoactivation
site, Sm target is irradiated together with $^{197}$Au which is used
as a reference. At the photon scattering site, the scattered photons
from $^{11}$B are observed using HPGe detectors. Another $^{197}$Au
reference target is sandwiched with $^{11}$B for flux calibration
purposes. }
\end{center}
\end{figure}

A schematic layout of the photoactivation facility is given in
Fig.~\ref{fig:experimentalsetup}. The primary electron beam is
focused onto a thin radiator foil which produces bremsstrahlung via
deceleration of electrons. The radiator is made of niobium with
areal densities varying between 1.7~mg/cm$^2$ and 10~mg/cm$^2$
corresponding to $1.6\times10^{-4}$ and $1\times10^{-3}$ radiation
lengths. Behind the radiator, the electrons are deflected by a
dipole magnet and dumped into a graphite cylinder with a conical
recess (see Fig.~\ref{fig:experimentalsetup},
$\textit{photoactivation site}$). The length of the cylinder is 600
mm and diameter is 200 mm. The photoactivation target (Sm) is
irradiated here together with the activation standard target
$^{197}$Au.

The bremsstrahlung beam goes straight ahead through the collimator
to the $\textit{photon scattering site}$. The photon-scattering site
is separated from the photoactivation site by a 1.6 m thick
heavy-concrete wall. The collimator made of high-purity aluminum is
placed 1 m behind the radiator. The collimator shapes a beam with
defined diameter from the spatial distribution of photons. For the
energy range of interest, the flux at the photon scattering site is
about 10$^{8}$ cm$^{-2}$ s$^{-1}$ MeV$^{-1}$. The endpoint energy of
the bremsstrahlung distribution is determined by measuring protons
from the photodisintegration of the deuteron (see
Fig.~\ref{fig:experimentalsetup}, deuteron breakup target) with
silicon detectors. From the maximum energy of the emitted protons,
the maximum energy of the incident photons was deduced. For a
detailed description of the beam energy determination procedure, see
Sec. III C of Ref.~\cite{Nair2008}.

At the photon scattering site, a $^{11}$B sample is irradiated
together with the activation standard target $^{197}$Au. The
activation standard reaction $^{197}$Au($\gamma,n$) has been
compared to photoneutron studies using monochromatic photons from
positron annihilation in flight technique (see \cite{Nair2008}, and
the references therein). The photon flux at this site is
experimentally determined by means of the known integrated cross
sections of the states in $^{11}$B depopulating via $\gamma$ rays.
The thin target bremsstrahlung and photon flux determination
procedure has been discussed in detail in Ref.~\cite{Nair2008}.

At the photoactivation site, the available photon flux is about 50
to 100 times higher compared to the photon scattering site. At this
high-flux area, it is technically not possible to determine the
photon flux online. The bremsstrahlung distribution by the graphite
is described by MCNP4C2~\cite{MCNP} simulations which are based on
the bremsstrahlung cross sections by Seltzer and
Berger~\cite{Selt86}.

For the activation measurements, the Sm/Au samples used were of
natural isotopic composition with masses $\sim$200 mg (Au) and
$\sim$2 g (Sm). The Sm powder was pressed into samples of cylindrical shape and
the masses were determined prior to the irradiation. After
irradiation, the decay of the daughter nuclei resulting from
photoactivation is studied with HPGe detectors of relative
efficiency 90\% or 60\% situated in a lead-shielded low-background
environment. Between the sample and the endcap of the detector, a
cadmium absorber with a thickness of 1.535 mm is inserted to block
the X-ray summing with the $\gamma$-rays. To maximize the absolute
efficiency of the detector, the source/target is put on top of the
Cd absorber which is situated directly on the HPGe capsule.

In case the resulting radioactive nuclei are short-lived, a rapid
transport system ($\textit{rabbit system}$) was used for activation
experiments. The rabbit system uses compressed air to transport the
samples between the photoactivation site and the measuring
site~\cite{Nair2007}. The samples to be irradiated are enclosed in
polyethylene cassettes and are shot to the high photon flux area
(photoactivation site, Fig.~\ref{fig:experimentalsetup}). After
irradiation, the samples are transported within about 15 s to the
detector for decay measurements. For the evaluation of fast decaying
samples, a list mode data acquisition system has been used which
enabled us to account for the rapidly changing dead time fractions
during the course of the measurement.

The measurements demanding low background were performed at the
underground laboratory 'Felsenkeller'~\cite{Koehl2009} in Dresden
where 98\% of the cosmic muons are shielded by a 47 m thick rock
layer. At Felsenkeller, the $\gamma$-decay was measured with a HPGe
detector of 30\% relative efficiency.

\section{Data Analysis}\label{sect:datanal}
The data analysis method is explained in the following sections. The
discussion is divided into two parts:
\begin{enumerate}[(A)]
\item the method of photoactivation and the photoactivation yield
determination; and
\item the photodisintegration reactions ($\gamma,n$),
($\gamma,p$) and ($\gamma,\alpha$) in $^{144}$Sm.
\end{enumerate}
The decay properties of the radioactive nuclei resulting from
photoactivation are given in Table ~\ref{tb:decayproperties}.

\begingroup
\begin{table*}[htb!] \caption{Decay properties of the daughter
nuclei stemming from the respective photodisintegration reactions in
$^{144}$Sm. The values are adopted from the Evaluated Nuclear
Structure Data Files of the National Nuclear Data Center online
service ~\cite{NNDC}.} \label{tb:decayproperties}
\begin{ruledtabular}
\begin{tabular}{clllll}
reaction & $S_{x}$\footnotemark[1] (MeV) & decay& $t_{1/2}$ \footnotemark[2] & $E_{\gamma}$\footnotemark[3] (keV)  & $\textit{p}$ \footnotemark[4] \\
\hline

\rule[0mm]{0mm}{5mm} $^{144}$Sm$(\gamma,n)$& 10.5200(24)
    &$^{143}$Sm(EC+$\beta^{+})$$^{143}$Pm & 8.75(8) min & 1056.58 & 0.019(2)\\
\rule[0mm]{0mm}{5mm}
    & & & & 1173.18 & 0.004161(544)\\
\rule[0mm]{0mm}{5mm}
    & & & & 1514.98 & 0.006593(75)\\
\rule[0mm]{0mm}{5mm}
    & & & & 1403.06 & 0.003496(414)\\
\rule[0mm]{0mm}{5mm}
    & &$^{143}$Sm IT decay & 66(2) s&  754.4 & 0.8988(6)\\
\hline
\rule[0mm]{0mm}{5mm} $^{144}$Sm$(\gamma,\alpha)$& -0.145 &$^{140}$Nd(EC)$^{140}$Pr(no $\gamma$) & 3.37(2) d &1596.11 &0.0050(4) \\
                                       & & $^{140}$Pr$(\beta^{+})$$^{140}$Ce
                                                          &  & & \\
\hline \rule[0mm]{0mm}{5mm} $^{144}$Sm$(\gamma,p)$& 6.295(3)
    &$^{143}$Pm(EC)$^{143}$Nd & 265(7) d & 741.98 & 0.385(24)\\
\end{tabular}
\end{ruledtabular}
\footnotetext[1]{separation energy for the proton or neutron
emission; Q-value for alpha emission} \footnotetext[2]{half-life of
the corresponding decay with absolute uncertainty given in
parentheses} \footnotetext[3]{$\gamma$-energy of the analyzed
transitions} \footnotetext[4]{photon emission probability per decay
with absolute uncertainty given in parentheses}
\end{table*}
\endgroup

\subsection{The photoactivation yield}
\label{sect:photoactivation-yield}

The photoactivation method essentially consists of two steps
\begin{enumerate}[(i)]
  \item{irradiation of the target nuclei ($^{144}$Sm
)}
  \item{measurement of decay of the daughter nuclei resulting from
  photoactivation}
\end{enumerate}
The irradiation is performed at the photoactivation site shown in
Fig.~\ref{fig:experimentalsetup}. The number of activated nuclei
$N_{\mathrm{act}}(E_0)$ produced is proportional to the integral of
the absolute photon fluence (time integrated flux)
$\Phi_{\gamma}(E,E_0)$ times the photodisintegration cross section
$\sigma _{(\gamma ,\mathrm{x})}(E)$. The integral runs from the
reaction threshold energy $E_{\mathrm{thr}}$ up to the endpoint
energy $E_0$ of the bremsstrahlung spectrum. The symbol $x$ = $n, p,
\alpha$ denotes the emitted particle. The number of target atoms in
the sample is denoted by $N_{\mathrm{tar}}$.
\begin{equation}
N_{\mathrm{act}}(E_0) =  N_{\mathrm{tar}} \cdot
\int_{E_{\mathrm{thr}}}^{E_{\mathrm{0}}} \sigma_{\mathrm{(\gamma
,x)}}(E)\cdot \Phi_{\gamma}(E,E_0)\,\rm{dE} \label{eqn:actnuclei}
\end{equation}
After irradiation, the $\gamma$-rays following the $\beta$-decays of
the radioactive sample are measured using HPGe detectors. The number
of activated nuclei $N_{\rm{act}}(E_0)$ is determined using the
relation:
\begin{equation}
N_{\mathrm{act}}(E_0) =
\frac{N_{\gamma}(E_{\gamma},E_0)\cdot\kappa_{\mathrm{corr}}}{\varepsilon
(E_{\gamma})\cdot p(E_\gamma)} \label{eqn:yint}
\end{equation}
$ N_{\gamma}(E_{\gamma},E_0)$, $\varepsilon(E_{\gamma})$ and
$p(E_\gamma)$ denote the dead-time and pile-up corrected full-energy
peak intensity of the observed transition, the absolute efficiency
of the detector at the energy $E_{\gamma}$ and the emission
probability of the photon with energy $E_\gamma$ respectively.

The factor $\kappa_{\mathrm{corr}}$ in Eq.~(\ref{eqn:yint}) is given
by
\begin{equation}
\kappa_{\mathrm{corr}}
=\frac{\exp{(\frac{t_{\mathrm{loss}}}{\tau})}}{1-\exp{(\frac{-t_{\mathrm{meas}}}{\tau})}}\cdot
\frac{\frac{t_{\mathrm{irr}}}{\tau}}{1-\exp{(\frac{-t_{\mathrm{irr}}}{\tau})}}
%{1-\exp{\frac{-t_{\mathrm{mess}}}{\tau}}}
\label{eqn:kappacorr}
\end{equation}
This expression determines the number of radioactive nuclei from
their decays measured during the time $t_{\mathrm{meas}}$. It also
takes into account decay losses during irradiation
($t_{\mathrm{irr}}$) and in between end of irradiation and beginning
of measurement ($t_{\mathrm{loss}}$). $\tau$ denotes the mean life
time of the radioactive nucleus formed during photoactivation.

For the $(\gamma,x$) reaction ($x=n,p$ \rm{or} $\alpha$), the
activation yield is denoted by $Y_{\mathrm{act}}$ and is expressed
as the ratio of the number of activated nuclei to the number of
target atoms in the sample. For example, for the
$^{144}$Sm$(\gamma,n)$ reaction,
\begin{equation}\label{eqn:yield}
Y_{\mathrm{act}}(\mathrm{^{143}\rm{Sm}}) =
\frac{N_{\mathrm{act}}(^{143}\rm{Sm})}{N_{\mathrm{tar}}(^{144}\rm{Sm})}
 \end{equation}
Using Eq.~(\ref{eqn:actnuclei}), the activation yield can be
calculated from $\sigma_{(\gamma ,n)}(E)$ data with the known
bremsstrahlung spectrum. In this way measured activation yields can
be compared with the experimental or theoretical cross section data.

To compare the activation yield measured at different endpoint
energies with the calculated yield using cross sections from
theory/previous experiments, the experimental data need to be
normalized to the photon fluence at the irradiation site (the
$\textit{photoactivation}$ $\textit{site}$, see
Fig.~\ref{fig:experimentalsetup}). As already mentioned, it is not
possible to measure the photon fluence at this site directly. The
fluence at a fixed energy $E_\gamma^\mathrm{X}$  is given by the
ratio of the measured $^{197}$Au($\gamma,n$) activation yield and
the calculated activation yield using the known $\sigma _{(\gamma
,n)}$ from $^{197}$Au and a simulated thick target bremsstrahlung
spectrum using the code MCNP. For the experiments discussed in this
paper, the normalization has been done to the fluence at
$E_\gamma^\mathrm{X}$ = 7.288 MeV which is well below the
bremsstrahlung endpoint energies for the photodisintegration
experiments under discussion.

\begin{figure}
\begin{center}
\includegraphics[width=7.5 cm,angle=270]{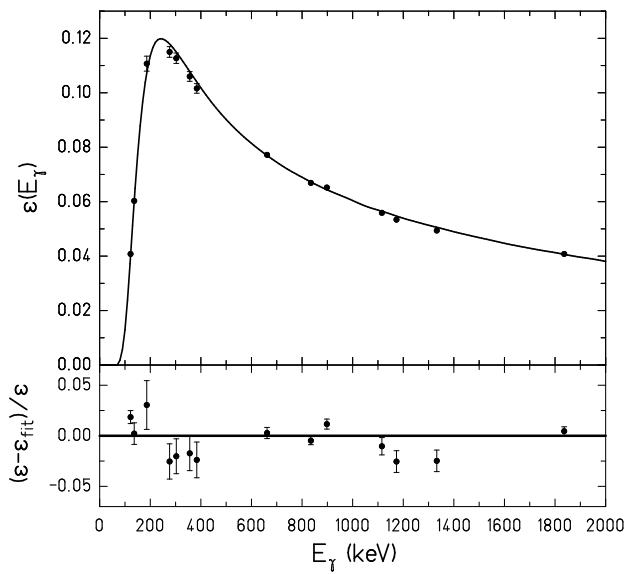}
%original:P:\Work\Data\calibration_LC-detectors\thesis-input\publication
\caption{\label{fig:LCeffic}Full-energy peak efficiency curve for a
HPGe detector with 90\% efficiency. The upper part is the fit curve
based on GEANT3 simulations normalized to the experimental data and
the lower part shows the relative residuals.}
\end{center}
\end{figure}

For the precise and accurate analysis of the decay spectra, the HPGe
detectors were calibrated with a certified set of calibration
sources in the energy range from 0.12 to 1.9 MeV. The efficiency
curve for the detector with 90\% relative efficiency is shown in
Fig.~\ref{fig:LCeffic}. For multi-gamma emitting nuclides,
coincidence summing corrections have been applied. The fit to the
data points is a curve based on a GEANT3~\cite{GEANT3} simulation
with realistic detector geometry. The detector dimensions have been
determined independently by performing a 200 keV X-ray scan. The
simulated and measured efficiencies agree within 1.01$\pm$0.02 for
the considered energy range. In the estimation of errors, both
statistical and systematic uncertainties are taken into account. The
systematic uncertainties in the activity as given in the source
certificates amount to 0.6-1.5\% in the energy range from 0.12 to
1.9 MeV. The statistical uncertainty mainly originates from the peak
fit for the calculation of peak areas and the dead-time and pile-up
corrections. The statistical uncertainty contribution is in the
order of 0.2-0.5\%. However, for the photoactivation experiment
described here, the targets used were not point-like and the
appropriate correction for volume-effects has been applied. This
will be described in the next section.

\subsection{Photodisintegration of $^{144}$Sm}

\begin{figure}
\begin{center}
\includegraphics[width=6 cm,angle=270]{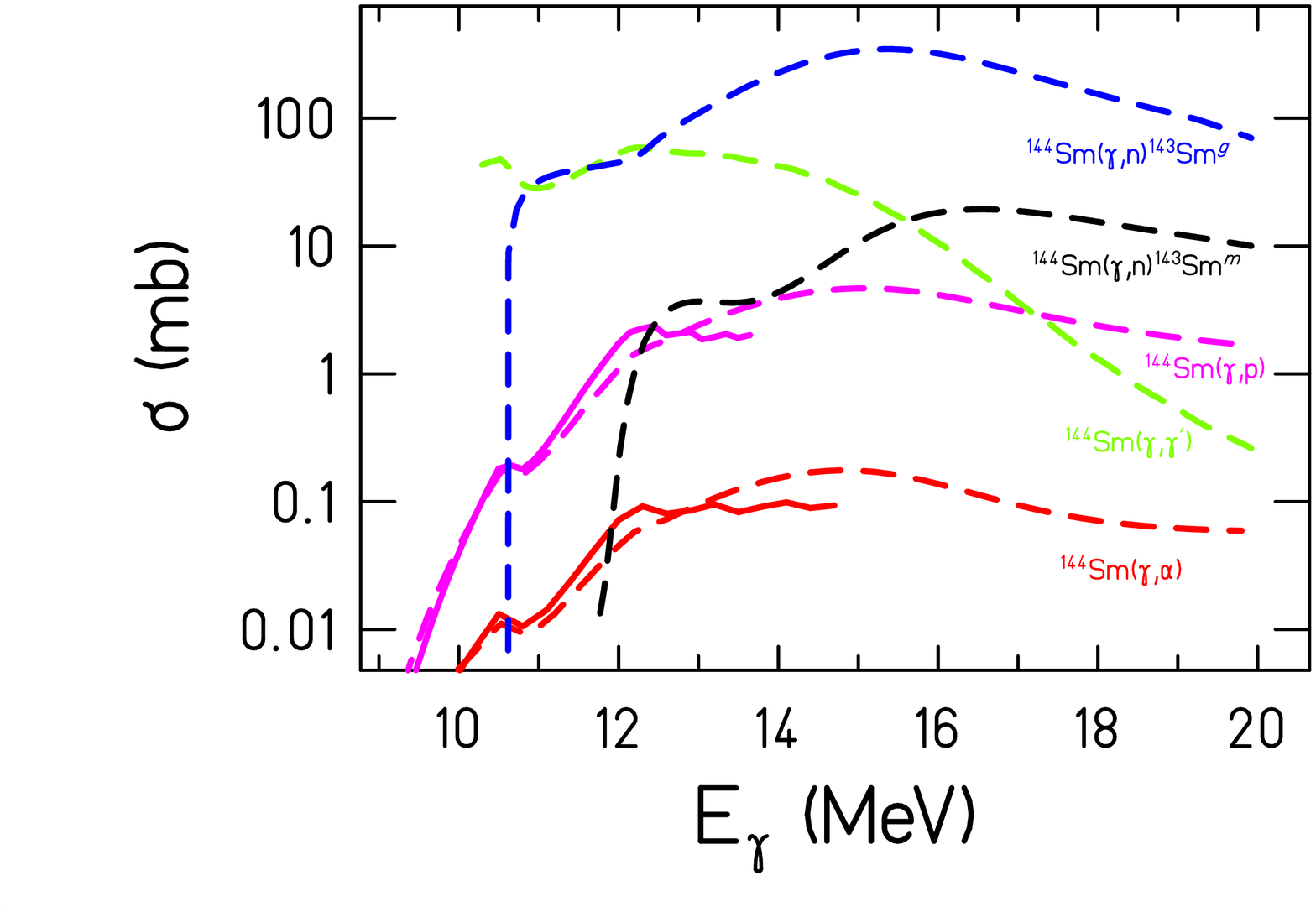}
%original:PP:\Work\Data\activation-Spectra\Sm144\publication-PRC
\caption {\label{fig:sm-gn-gp-ga-crosssec}(Color online)
$^{144}$Sm$(\gamma,\gamma')$ (green),$^{144}$Sm$(\gamma,n)$ (blue to the ground state, black to the isomer),
$^{144}$Sm$(\gamma,p)$ (purple) and $^{144}$Sm$(\gamma,\alpha)$ (red) cross
sections from model calculations. TALYS and NON-SMOKER predictions
are shown with dashed and solid lines respectively. The TALYS
calculations were performed with the default input parameter set in
the version TALYS-1.0.}
\end{center}
\end{figure}

In this experiment, the photodisintegration reactions ($\gamma,n$),
($\gamma,p$) and ($\gamma,\alpha$) on $^{144}$Sm were studied. The
cross section predictions for the ($\gamma,n$), $(\gamma,p)$ and
$(\gamma,\alpha)$ reactions by the statistical models
TALYS~\cite{Koni05} and NON-SMOKER~\cite{Raus04} are given in
Fig.~\ref{fig:sm-gn-gp-ga-crosssec}. The $^{144}$Sm$(\gamma,n)$
reaction produces $^{143}$Sm or $^{143\rm{m}}$Sm. The NON-SMOKER
model doesn't provide the cross section data for the ground and
isomeric states in $^{144}$Sm separately.  Above 11 MeV, the
($\gamma,p$) and ($\gamma,\alpha$) cross sections are orders of
magnitude lower than the ($\gamma,n$) cross sections due to the
suppressive effect of the Coulomb barrier. The reaction thresholds
and decay properties of the respective photodisintegrations in
$^{144}$Sm are given in Table~\ref{tb:decayproperties}.

For studying the $^{144}$Sm$(\gamma,n)$ reaction, the irradiation
was performed for various endpoint energies starting with E$_{0}$
= 11.00 MeV which is approximately 480 keV above the neutron
separation energy. The daughter nucleus produced is $^{143}$Sm,
partially in its isomeric state $^{143}$Sm$^{\it{m}}$. The
$^{144}$Sm($\gamma,n$)$^{143}$Sm$^{\it{m}}$ reaction is identified
with the isomeric transition at 754 keV and the
$^{144}$Sm($\gamma,n$)$^{143}$Sm$^{\it{g}}$ with transitions above
1000 keV (see Table~\ref{tb:decayproperties}). Since both
radionuclides are short-lived, the irradiation was carried out
using the rabbit system. The activations for this experiment
lasted between 15-45 minutes. The sample spectrum of an irradiated
Sm target and the half-life measurement using the exponential
decay of the ground state of $^{143}$Sm have been presented in
Fig. 2 of Ref.~\cite{Nair2007}.

For the activation using the rabbit system, the targets used were
not point-like. For the irradiation of samarium, we used fine
Sm$_2$O$_3$ powder filled in a cylinder (length 16 mm, diameter 10
mm) of about 2-3 g mass. The $^{197}$Au activation standard targets
were made of Au-metal foils of about 200-250 mg mass rolled into a
cylindrical shape (length 36 mm, diameter 5 mm). Therefore, the
full-energy peak efficiencies measured with point sources have to be
corrected for source-extension and self-absorption effects. To
account for the volume source effect, detailed simulations using
GEANT3 were performed. The absolute full-energy peak efficiency of
the target is given by
\begin{eqnarray}
\label{eqn:rp-photopeakeff} \varepsilon_{\textrm
{target}}\left(E_{\gamma}\right) = \varepsilon_{\textrm
{point}}\left(E_{\gamma}\right) \cdot
\frac{\varepsilon^{\textrm{sim}}_{\textrm{target}}\left(E_{\gamma}\right)}{\varepsilon^{\textrm{sim}}_{\textrm{point}}\left(E_{\gamma}\right)}
\end{eqnarray}

$\varepsilon_{\textrm {point}}$ denotes the efficiencies measured
with point-like sources at a distance corresponding to the distance
of the center of the volume source to the HPGe crystal. The
simulated full-energy peak efficiencies for volume and point-like
sources are denoted by
$\varepsilon^{\textrm{sim}}_{\textrm{target}}$ and
$\varepsilon^{\textrm{sim}}_{\textrm{point}}$ respectively. The
ratio is the correction to be applied for the volume source effect.
For example, for the transition at 754 keV resulting from the
$^{143\rm{m}}$Sm decay, the correction factor
$\left(\frac{\varepsilon^{\textrm{sim}}_{\textrm{target}}\left(E_{\gamma}\right)}{\varepsilon^{\textrm{sim}}_{\textrm{point}}\left(E_{\gamma}\right)}\right)$
amounts to about 0.97$\pm$0.03. For the transition at 356 keV
following the decay of $^{196}$Au, the correction factor was about
0.93$\pm$0.02.

\begin{figure}[htb!]
\begin{center}
\includegraphics[width=6 cm,angle=270]{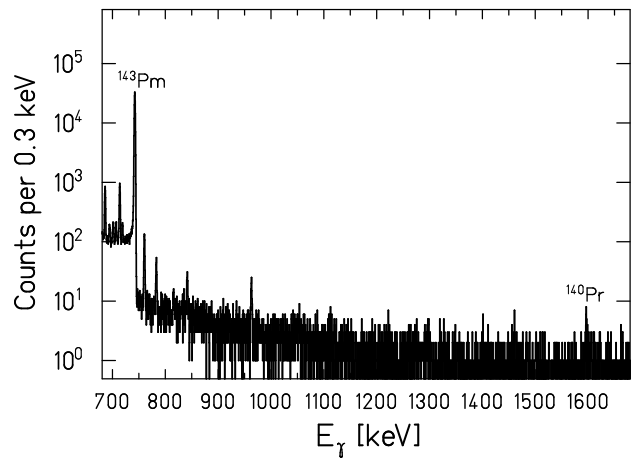}
% figure location :P:\Work\Data\activation-Spectra\activation-2006-2007\Felsenkeller-August2007\samarium-target\publication
\end{center}
\caption{\label{fig:sm-gp-ga-peaks}The spectrum of an irradiated
Sm$_{2}$O$_{3}$ target in which the decays following the
$^{144}$Sm$(\gamma,p)$ and $^{144}$Sm$(\gamma,\alpha)$
disintegrations are observed. The strongest $\gamma$-transitions
following the $\beta$-decays of $^{143}$Pm and $^{140}$Pr are
marked. The target was placed on top of the 30\% HPGe detector at
the underground laboratory. The other peaks are from the
contamination reactions like $^{154}$Sm$(\gamma,n)$ etc. The target
was irradiated at a bremsstrahlung endpoint energy of 15.0 MeV and
the spectrum shown here was recorded for a period of 1 day.}
\end{figure}

The $^{144}$Sm$(\gamma,p)$ and $^{144}$Sm$(\gamma,\alpha)$
activation yields were measured for the first time at
astrophysically relevant energies. In the context of this work,
these are categorized as $\textit{manual}$ measurements, indicating
that the rabbit system was not necessary for the transport of the
rather long-lived isotopes to the counting setup.

For the energy range considered in these experiments, the
$^{144}$Sm$(\gamma,p)$ and $(\gamma,\alpha)$ cross sections are
rather very small compared to the $(\gamma,n)$ cross sections (see
Fig.~\ref{fig:sm-gn-gp-ga-crosssec}). This results in a decay
spectrum with weak counting statistics. Therefore, the decay
measurements were performed under optimized background conditions
in the 'Felsenkeller' underground laboratory.

In the photoactivation of $^{144}$Sm, the $(\gamma,n)$ and
$(\gamma,p)$ reaction yields cannot be distinguished at high
energies above the neutron separation energy ($S_{n}=10.52$ MeV)
since the $(\gamma,n)$ daughter nucleus decays very quickly to the
$(\gamma,p)$ daughter nucleus. Therefore, to measure the pure
$(\gamma,p)$ reactions, we used bremsstrahlung energies
sufficiently below 10.52 MeV.

A sample spectrum of an irradiated samarium target is given in
Fig.~\ref{fig:sm-gp-ga-peaks}. The Sm$_{2}$O$_{3}$ target was
placed on top of the 30\% HPGe detector. In this measurement, the
target was irradiated at a bremsstrahlung endpoint energy of 15.0
MeV and was brought to the underground laboratory on the next day
of irradiation. The spectrum shown here was recorded for a period
of 1 day. The $^{144}$Sm$(\gamma,p)$ reaction is identified by the
transition at 742 keV from $^{143}$Pm decay (EC) with a half-life
of 265 days, see figure. The $^{144}$Sm($\gamma$,$\alpha$)
reaction was identified by the transition at 1596 keV following
the decay of $^{140}$Pr which is the daughter of the
($\gamma$,$\alpha$) product $^{140}$Nd. In the spectrum shown in
Fig.~\ref{fig:sm-gp-ga-peaks}, the 1596 keV transition is also
marked. The other peaks in the spectrum stem from the
contaminating reactions in the other samarium isotopes (e.g.,
$^{154}$Sm$(\gamma,n)$).

For the $\textit{manual}$ measurements, we used Sm$_{2}$O$_{3}$
powder filled into discs of radius 9 mm and thickness 5 mm. To
correct the full-energy peak efficiencies for the volume source
effect, we used the Monte-Carlo efficiency transfer code
EFFTRAN~\cite{Vidm05}. The point source efficiency has been measured
with the point source shifted to a distance which corresponds to the
center of the volume source. The point source efficiency is reduced
by 7\% when the point source is shifted and put in  the middle of
the volume target. The volume source effect is only 4\%, comparing
the point source at the center of the Sm$_{2}$O$_{3}$ sample and a
volume target Sm$_{2}$O$_{3}$ placed on the top of the cadmium
absorber. The $^{197}$Au activation standard targets were very thin
discs (mass$\sim$100 mg, thickness-0.2 mm, radius-9 mm) and the
volume source effect was not significant.

The activation yield relative to the photon fluence are presented in
the next sections in detail.
\section{Results and Discussion}
\label{sect:sm-yield-results}
\subsection{Activation yield for $^{144}$Sm$(\gamma,n)$ reaction: Comparison to previous experiments}
\label{sect:sm-gnyield}

\begin{figure}[htb!]
\begin{center}
\includegraphics[width=6 cm, angle=270]{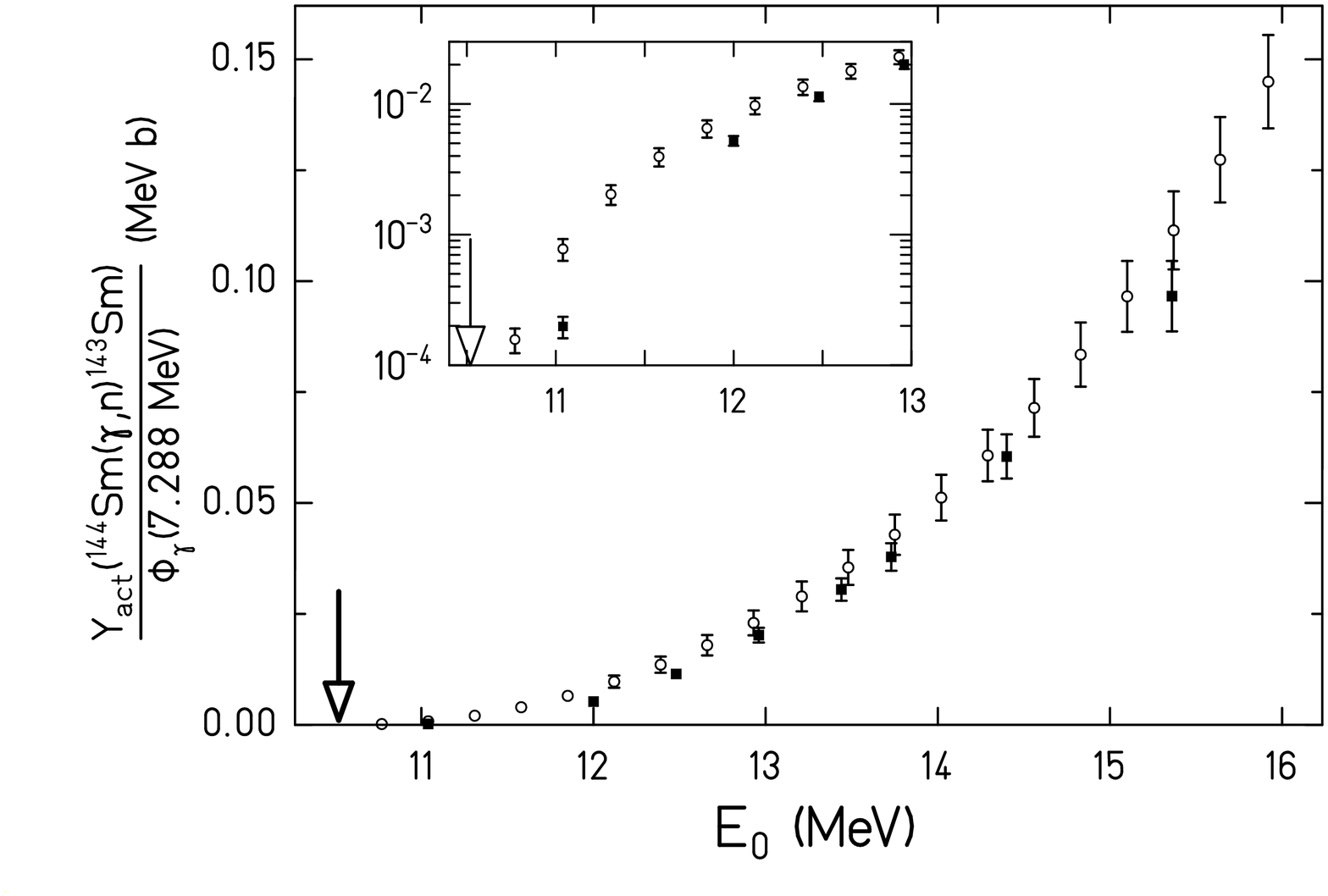}
\end{center}
%location: P:\Work\Data\activation-Spectra\rohrpost\RPact_Oct06\publication-PRC\yield
\caption{\label{fig:sm-gn-yield-gnd}The experimental activation
yield from ELBE (squares) for the $^{144}$Sm$(\gamma,n)$$^{143}$Sm
reaction normalized to the photon fluence is compared to the yield
calculated using cross sections measured by Carlos $\textit{et al}$.
(open circles). The data from Carlos $\textit{et al}$.~\cite{Carl74}
have been scaled by factor of 0.8, following the recommendations by
Berman $\textit{et al}$.~\cite{Berm87} (see text). The downward
arrow denotes the neutron emission threshold.}
\end{figure}

The activation yield for the $^{144}$Sm($\gamma,n$) reaction was
determined using the method discussed in
Sec.~\ref{sect:photoactivation-yield}. The experimental activation
yield for the $^{144}$Sm($\gamma$,n)$^{143}$Sm reaction ($Y_{g}$)
relative to the photon fluence is given in
Fig.~\ref{fig:sm-gn-yield-gnd}.

The activation yield is compared to the yield calculated using cross
sections from previous experiments. The photoneutron cross sections
of the Sm isotopes have been measured by Carlos $\textit{et
al}$.~\cite{Carl74} using positron annihilation in flight beams at
Saclay. In particular, Carlos $\textit{et al}$. present a study of
the transition from spherical to deformed shape for isotopes in the
samarium region. The partial photoneutron cross sections -
[$\sigma$($\gamma$,n)+$\sigma$($\gamma$,np)] and
$\sigma$($\gamma$,2n) of $^{144}$Sm are given in Fig.~2 of
\cite{Carl74}.

\begin{figure}[htb!]
\begin{center}
\includegraphics[width=6 cm, angle=270]{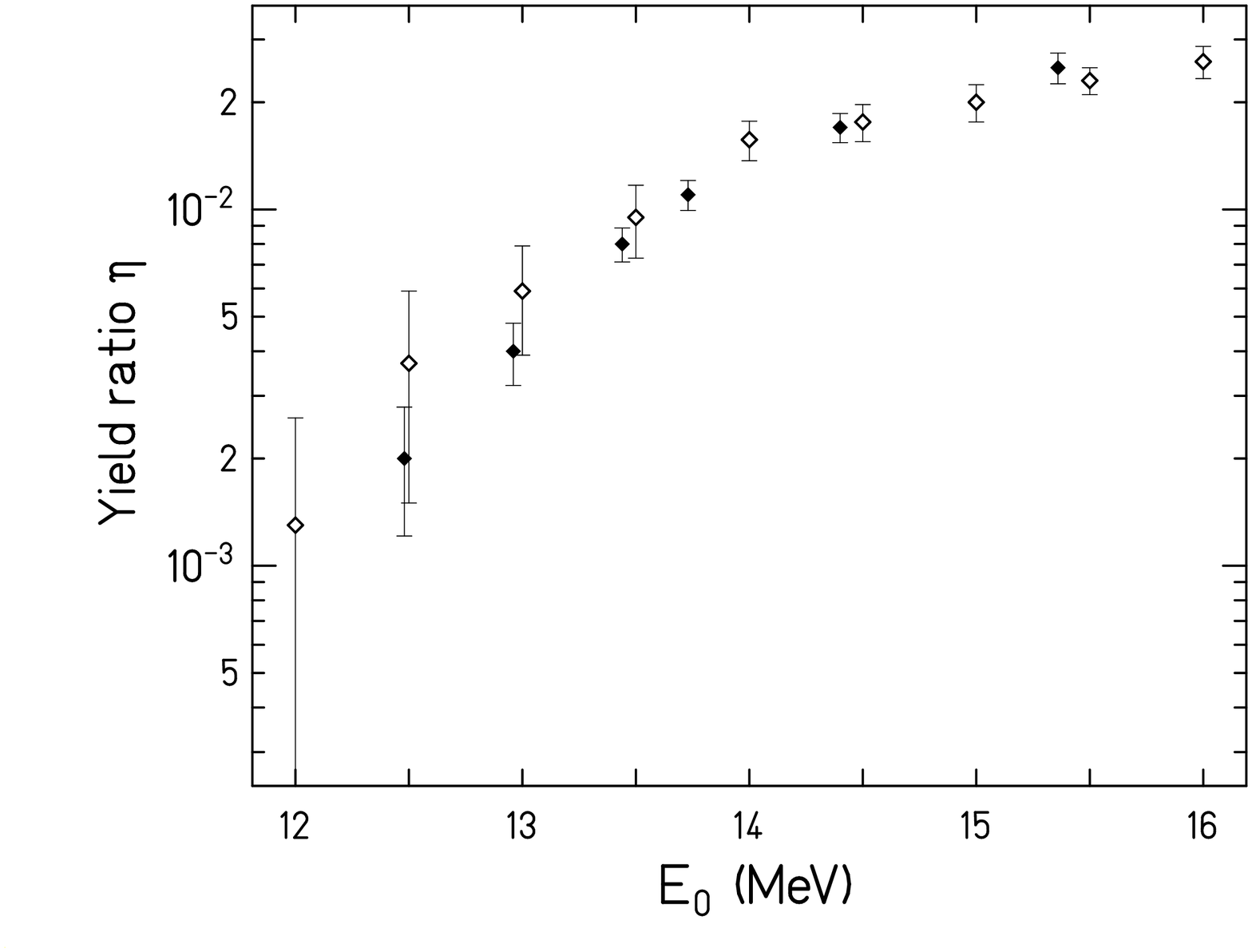}
\end{center}
%location: P:\Work\Data\activation-Spectra\rohrpost\RPact_Oct06\publication-PRC\yield
\caption{\label{fig:sm-gn-yield-isom}The isomeric yield ratio $\eta
= \frac {Y_{m}} {Y_{m}+Y_{g}}$ measured at ELBE (diamonds) for the
$^{144}$Sm($\gamma$,n) reaction is compared to the isomeric yield
ratio given by Mazur $\textit{et al}$. (open diamonds).}
\end{figure}

For calculating the activation yield, the cross section data from
Carlos $\textit{et al}$. were renormalized by us to correct for two
effects:
\begin{enumerate}
\item The target used by Carlos $\textit{et al}$. had a contamination of 11.4\% of other Sm-isotopes. We have
subtracted the possible contamination using earlier data obtained
with natural Sm at the same laboratory~\cite{Berg69} at the same
photon energies. For the energy range 11-16 MeV, the correction
factor for the isotopic contamination is 0.8 to 0.9.
\item As shown in a precision experiment by Berman $\textit{et al}$.~\cite{Berm87}, the cross sections measured at Saclay for nuclei with $A<150$ are
larger by approximately 20\% in all cases except $^{141}$Pr for
which a comparison was made. For this particular case, an additional
correction factor of 0.8 was applied.
\end{enumerate}
The corrected cross sections were then used to calculate the
activation yield which compares to our experimental yield as shown
in Fig.~\ref{fig:sm-gn-yield-gnd}.

In addition, the activation yield $Y_{m}$ for the
$^{144}$Sm($\gamma$,n)$^{143m}$Sm reaction has been determined using
the method described in Sec.~\ref{sect:photoactivation-yield}. With
known $Y_{g}$ and $Y_{m}$, the isomeric yield ratio $\eta = \frac
{Y_{m}} {Y_{m}+Y_{g}}$ was determined (see
Fig.~\ref{fig:sm-gn-yield-isom}).

In an activation experiment by Mazur $\textit{et al}$.~\cite{Mazu95}
using bremsstrahlung beams from the M-30 microtron facility at
Uzhgorod (Ukraine), the isomeric-state excitations have been
investigated for several $N=82$ closed-shell nuclei  in the energy
range 8-18 MeV. The $(\gamma,n)^{m}$ cross sections were deduced
using the Penfold-Leiss method~\cite{Penf59}, see Fig.~1 of
Ref.~\cite{Mazu95}. Mazur $\textit{et al}$. have also measured the
isomeric yield ratio $\eta = \frac {Y_{m}}
{Y_{m}+Y_{g}}$~\cite{Vans81} as a function of energy for the
$^{144}$Sm($\gamma$,n)$^{143g,m}$Sm reaction. The experimental
isomeric yield ratio from the ELBE is in good agreement with the
data from Mazur $\textit{et al}$, see
Fig.~\ref{fig:sm-gn-yield-isom}. At low energies close to the
reaction threshold, the activation yield is influenced by the
different experimental conditions (e.g, use of a hardener,
thin/thick radiators etc).

\subsection{Model calculations in the Hauser-Feshbach formalism}
\label{sect:sm-gnyield-modelcalc}

This section is dedicated to the comparison of experimental results
to the predictions of the two advanced model codes TALYS and
NON-SMOKER which are based on the Hauser-Feshbach formalism. The
NON-SMOKER code uses the neutron optical-model potential by Jeukenne
$\textit{et al}$.~\cite{Jeuk77} with a low-energy modification by
Lejeune~\cite{Leje80}. The $\gamma$-ray strength function is taken
from Thielemann and Arnould~\cite{Thie83}. The low-energy
modification of the GDR Lorentzian is by McCullagh $\textit{et
al}$.~\cite{McCu81}. The nuclear level density implemented in the
NON-SMOKER code is based on a global parametrization by Rauscher
$\textit{et al}$.~\cite{Raus97} within the back-shifted Fermi-gas
formalism.

The TALYS calculations presented in this paper were performed with
the current version of the code TALYS-1.0. The default option of this
code uses the neutron optical-model potential parameterizations by
Koning and Delaroche~\cite{Koni03}. The E1 photon strength function
is from the compilation by Kopecky and Uhl~\cite{Kope90}. The
nuclear level density model is based on an approach using the
Fermi-gas model~\cite{Koni08}.

\begin{figure}[htb!]
\begin{center}
\includegraphics[width=6 cm, angle=270]{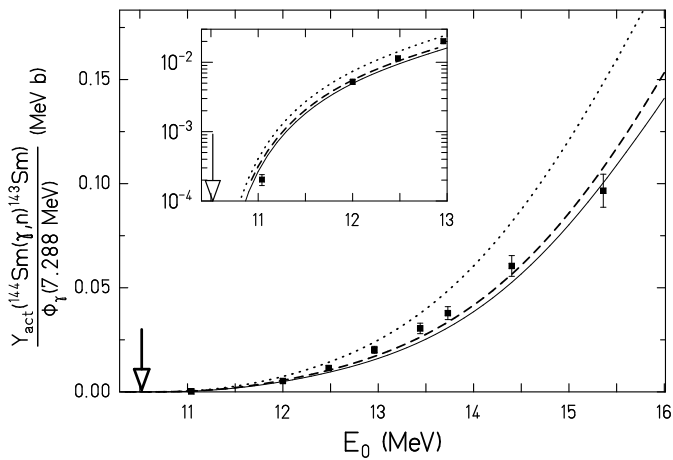}
\end{center}
%location: P:\Work\Data\activation-Spectra\rohrpost\RPact_Oct06\publication-PRC\yield
\caption[$^{144}$Sm$(\gamma,n)$$^{143}$Sm activation yield compared
to the yield calculated using cross sections from model
calculations]{\label{fig:sm-gn-yield-theory-gnd}Experimental
activation yield relative to the photon fluence for the
$^{144}$Sm$(\gamma,n)$ reaction compared to theoretical model
calculations. The experimental data are denoted by squares with a
downward arrow denoting the neutron emission threshold. The dashed
and dotted lines denote yield calculations using cross sections from
TALYS and NON-SMOKER codes respectively. The solid line represents a
TALYS calculation with modified inputs, see text.}
\end{figure}

The measured activation yield ($Y_{g}$) is compared to the simulated
yield from  Hauser-Feshbach models as shown in
Fig.~\ref{fig:sm-gn-yield-theory-gnd}. It is observed that the
experimental yield roughly agrees to the simulated yield using cross
sections predicted by the default inputs to TALYS (agrees within
20\%) and NON-SMOKER (agrees within a factor of 2).

The statistical model calculations are sensitive to the basic input
ingredients like optical potentials, strength functions and level
densities. In the case of ($\gamma$,n) reactions, the influence of
the neutron optical potential is weak, whereas the photon strength
function is a crucial ingredient of the model calculations. We
modified the deformation dependent parameters of the E1 strength
function used in TALYS according to a new parametrization explained
below (see also Ref.~\cite{Junghans2008}).

The strength function $f_{1}(E_{\gamma})$ is related to the average
photoabsorption cross section
$\overline{\sigma_{\gamma}}(E_{\gamma})$ over a large number of
levels with same spin and parity by the equation
\begin{equation}\frac{2J_{0}+1}{2J_{x}+1}\cdot
\frac{\overline{\sigma_{\gamma}}(E_{\gamma})}{(\pi\hbar\mathrm{c})^2
E_{\gamma}} = f_{1}(E_{\gamma}) \label{eq:sigma}
\end{equation}
$J_{0}$ and $J_{x}$ denote the spins of ground and excited states
respectively~\cite{Bartholomew1973}. Eq.~\ref{eq:sigma} can be
applied to a wide energy range around the GDR, including the low
energy tail of it. In even nuclei, the spin $J_{x}$ is equal to the
multipolarity of the GDR, i.e., $\lambda$ = 1.

In nearly all types of heavy nuclei, the GDR can be treated as
splitted into two or three components, whose energies are well
predicted by the finite range droplet model (FRDM)~\cite{Myers1977}.
Making use of the fact that the vibrational frequency $E_{k}/\hbar$
along a given axis $k$ is inversely proportional to the
corresponding semi-axis length $R_{k}$, and the
splitting~\cite{Bush1991} is due to the three different axes of the
ellipsoid, the nuclear shape with its quadrupole deformation
parameter $\beta$ and triaxiality parameter $\gamma$ is parameterized
as:
\begin{equation}
E_{k} = \frac{E_{0}\cdot R_{0}}{R_{k}} =
\frac{E_{0}}{\exp\left[\sqrt{\frac{5}{4 \pi}}\cdot \beta
\cdot\mathrm{cos}(\gamma-\frac{2}{3}k\pi)\right]} \label{eq:GDR}
\end{equation}
Here, $R_{0}$ denotes the nuclear radius and $E_{0}$ is the GDR
centroid energy. For a spherical nucleus with mass $A$, $R_{0}$ is
given by $R_{0}=1.16 A^{1/3}$ fm and $E_{0}$ is calculated with an
effective nucleon mass $m^*$= 874 MeV/c$^2$.

In the proposed phenomenological ansatz we parameterize the strength
function by a sum of three Lorentzians tailing down from the GDR.
The centroid energy of the GDR is directly related to parameters of
the finite range droplet model with one extra parameter which is
equal for all heavy nuclei, the reduced nucleon mass
m*~\cite{Moll95}. As was shown theoretically~\cite{Dover1972} a
description of the $E1$ photoabsorption cross section by a
Lorentzian is appropriate although the total width $\Gamma$ of a GDR
in a heavy nucleus is dominated by spreading and not by escape, i.e.
direct decay.

The average absorption cross section in the GDR is given by
\begin{equation}
\overline{\sigma_{\gamma}}(E_{\gamma}) = \frac{1.29\cdot Z\cdot
N}{A}\sum_{\rm
k=1}^3\frac{E_{\gamma}^2\Gamma_{k}}{(E_{k}^2-E_{\gamma}^2)^2+E_{\gamma}^2\Gamma_{k}^2}
\label{eq:Gamma}
\end{equation}
Here we assume that the GDR width $\Gamma_{k}$ to be used in the sum
of the three Lorentzians depends on the resonance energy $E_{k}$
only, in contrast to earlier descriptions~\cite{Kope90,Zanini2003}.
The photon energy $E_\gamma$ is given in MeV and
$\sigma_{\gamma}(E_{\gamma})$ in fm$^2$.

In this description, the Thomas-Reiche-Kuhn sum rule as determined
from general quantum mechanical arguments~\cite{Eisen88} is included
for obtaining the average photoabsorption cross section on an
absolute scale. We use results from hydrodynamical
considerations~\cite{Bush1991} by adapting surface dissipation to
the Goldhaber-Teller model of the GDR. This results in a power law
dependence of the resonance width on the respective resonance energy
$E_k$. In general, for all stable nuclei with $A>80$,
\begin{equation}
\Gamma_{k}(E_{k}) = 0.05\cdot \left(E_{k}\right)^\delta ,
\label{eq:gammak}
\end{equation}
For the exponent, the value $\delta$ = 1.6 was derived from the
one-body dissipation model~\cite{Bush1991} and the proportionality
parameter stems from a fit to many nuclei including axially as well
as triaxially deformed nuclei where the GDR is split into three
components with well defined distances from the centroid and
relative intensities. We generalize this by using the
proposed~\cite{Bush1991} power law dependence for different nuclei
and thus reduce the number of parameters for the description of the
GDR's by relating the spreading width in all nuclei with mass $A>80$
to the respective resonance energy of their GDR. It is stressed here
again that the Lorentzian photon strength function $f(E_{r})$ is
parameterized in each nucleus by a constant spreading width
$\Gamma(E_{k})$ $\textit{not}$ depending on the photon energy.

The above parametrization has been discussed in detail with examples
from experimental studies of 8 nuclides between $A$ = 80 and 238 in
a recent publication~\cite{Junghans2008}. The new phenomenological
description based on the ground state deformation parameters
describes well the electric dipole excitations for nuclei with $A
>80$ from $E_{x}\approx$ 4 MeV up to several MeV above the GDR. In
addition, the dipole strength function for the case of $^{197}$Au
was determined using the proposed parametrization and compared with
experimental results from ELBE~\cite{Nair2008}.

\begin{figure}[htb!]
\begin{center}
\includegraphics[width=6 cm, angle=270]{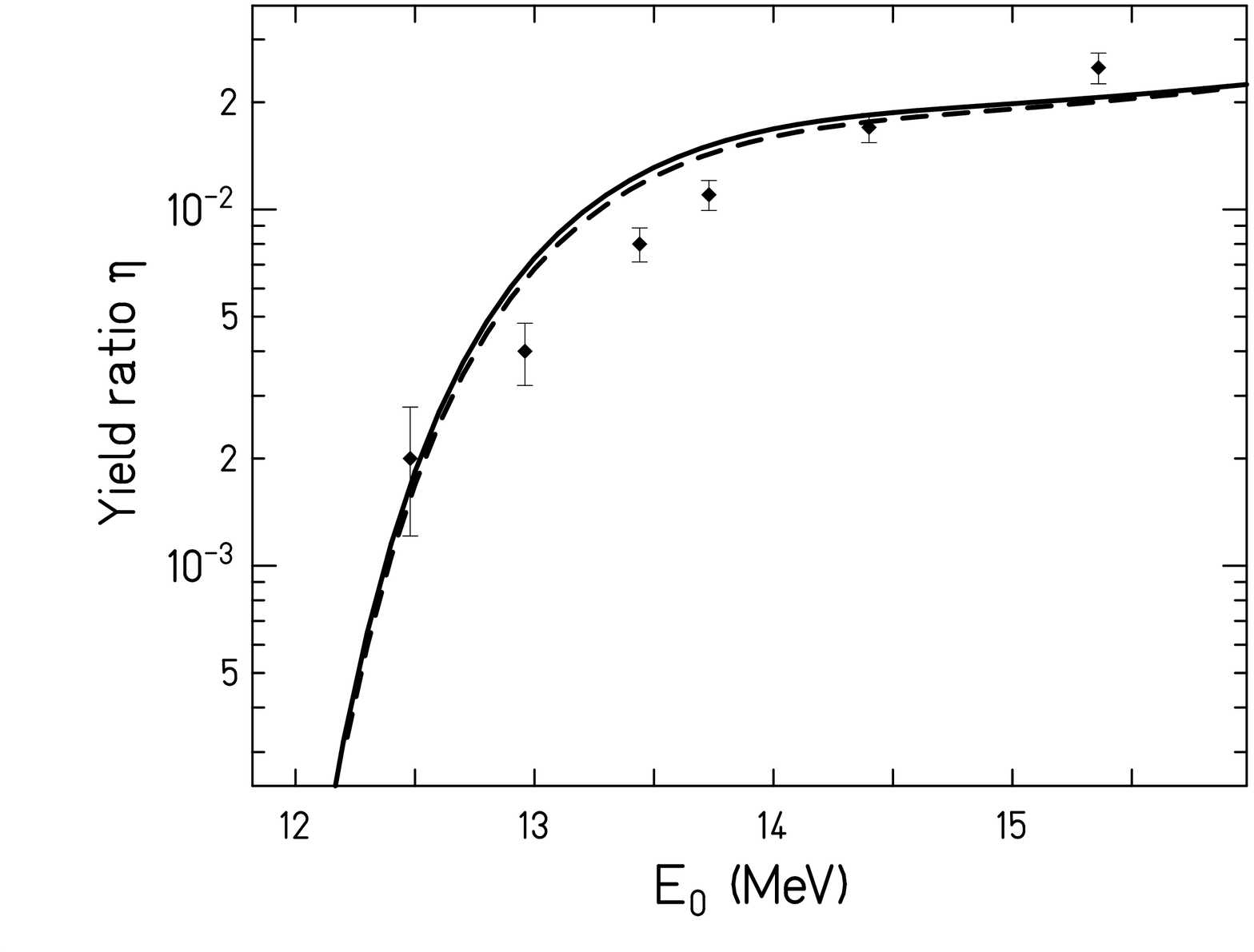}
\end{center}
%location: P:\Work\Data\activation-Spectra\rohrpost\RPact_Oct06\publication-PRC\yield
\caption[The isomeric yield ratio measured at ELBE for the
$^{144}$Sm($\gamma$,n) reaction compared to the simulated isomeric
yield ratio calculated using cross sections from model
calculations]{\label{fig:sm-gn-yield-theory-isom}The isomeric yield
ratio $\eta = \frac {Y_{m}} {Y_{m}+Y_{g}}$ measured at ELBE for the
$^{144}$Sm($\gamma$,n) reaction (diamonds) is compared to the
isomeric yield ratio calculated using cross sections predicted by
the TALYS code (dashed line). The solid line represents the yield
ratio calculated using modified inputs to the photon strength
functions, see text.}
\end{figure}

The yield for the $^{144}$Sm$(\gamma,n)$ reaction calculated using
the TALYS code with modified E1 photon strength functions derived on
the basis of the above discussion is shown in
Fig.~\ref{fig:sm-gn-yield-theory-gnd}. As in the quasi-spherical
$^{144}$Sm the three GDR components coincide, the splitting is
negligible and the following resonance parameters were used: $E_{0}$
= 14.84 MeV, $\sigma_{0}$ = 356.6 mb and $\Gamma_{0}$ = 3.74 MeV. In
Fig.~\ref{fig:sm-gn-yield-theory-isom}, the isomeric yield ratio
$\eta$ measured at ELBE is compared to the isomeric yield ratio
calculated using cross sections predicted by the TALYS code with
default and modified inputs. The NON-SMOKER model does not provide
the cross sections arising from the decay of the metastable state in
$^{143}$Sm separately and hence it was not possible to calculate a
yield ratio.

\subsection{Activation yield for $^{144}$Sm$(\gamma,p)$ and $^{144}$Sm$(\gamma,\alpha)$ reactions}
\begin{figure}[htb!]
\begin{center}
\includegraphics[width=6 cm,angle=270]{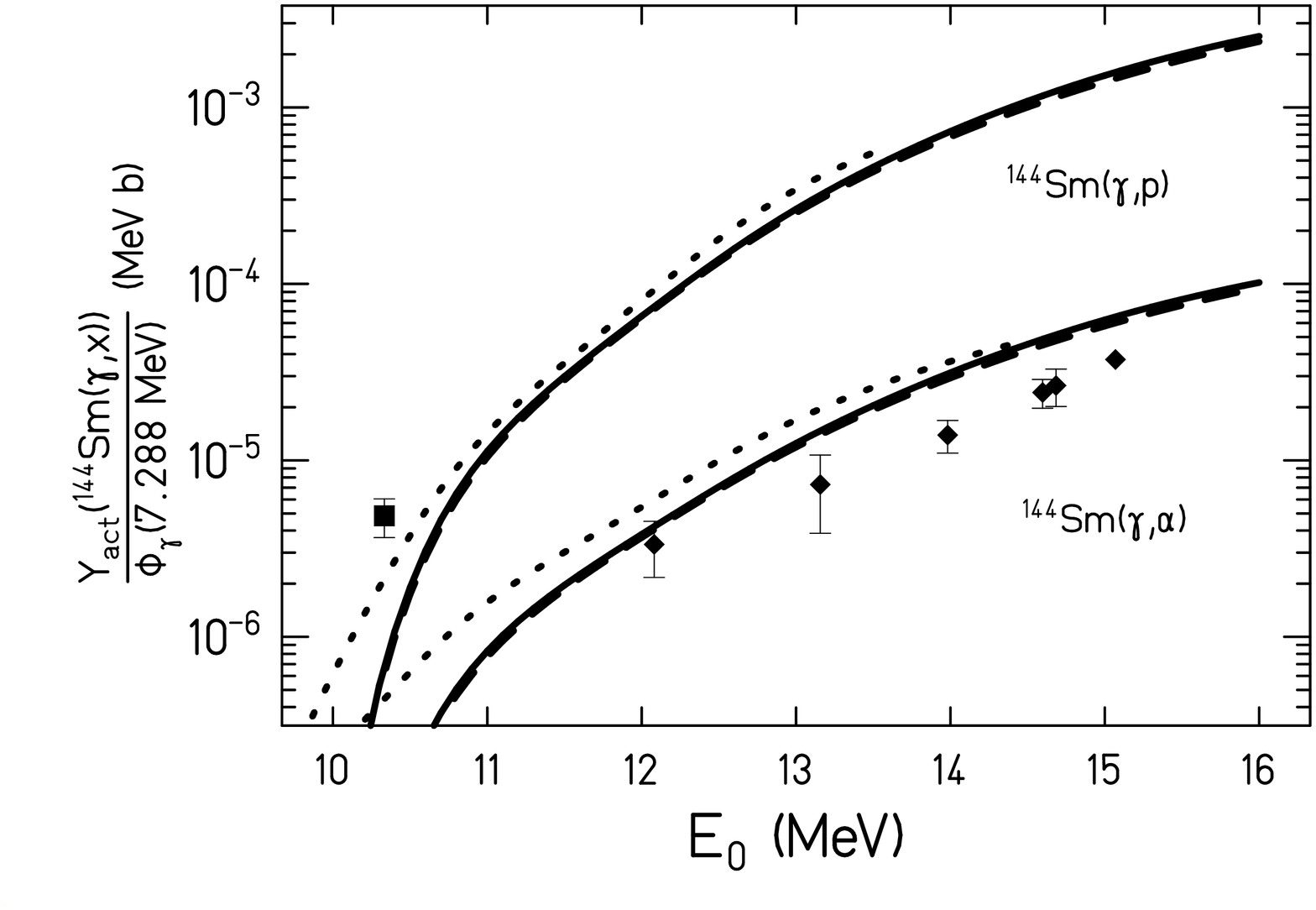}
%location:P:\Work\Data\activation-Spectra\activation-2006-2007\Felsenkeller-August2007\samarium-target\publication
\end{center}
\caption{\label{fig:sm-gp-ga-yield}Experimental activation yield
relative to the photon fluence for the ($\gamma$,p) and
($\gamma,\alpha$) reactions in $^{144}$Sm. The yields for $^{144}$Sm
($\gamma$,p) (squares) and $^{144}$Sm($\gamma$,$\alpha$) (diamonds)
are shown. The dashed and dotted lines denote yield calculations
using cross sections from TALYS and NON-SMOKER codes with default
inputs whereas the solid line represents a TALYS calculation with
modified inputs, see text.}
\end{figure}

\begin{figure*}[htb!]
\begin{center}
\includegraphics[width=0.5\textwidth,angle=270]{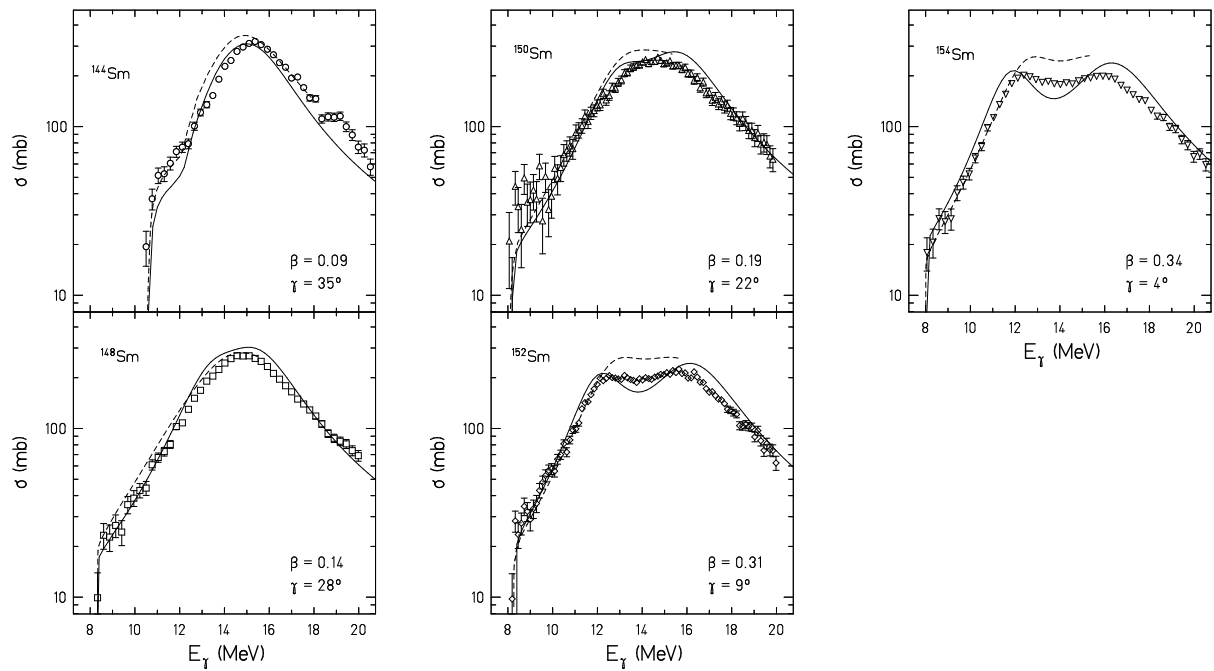}
%original:P:\Work\Data\activation-Spectra\rohrpost\RPact_Oct06\publication-PRC\Carlos-EG-photonstr
\caption {\label{fig:sm-gx-parametrization}The photoneutron cross
sections $\sigma$ [$\sigma$($\gamma$,n)+$\sigma$($\gamma$,np)] for
Sm-isotopes as described by the photon strength parametrization
(solid lines) from Junghans $\textit{et al}$.~\cite{Junghans2008}.
The original cross sections from Carlos $\textit{et
al}$.~\cite{Carl74} have been scaled by a correction factor of
0.8, see text. The dashed lines denote yield calculations using
cross sections from the NON-SMOKER code with default nuclear
physics inputs.}
\end{center}
\end{figure*}

The activation yields for the $^{144}$Sm$(\gamma,p)$ and
$^{144}$Sm$(\gamma,\alpha)$ reactions compared to the simulated
yields using cross sections predicted by the TALYS and NON-SMOKER
calculations are shown in Fig.~\ref{fig:sm-gp-ga-yield}. The
predictions given by the TALYS code with modified inputs to the E1
photon strength function is also shown.

The uncertainties in the experimental yield shown in
Fig.~\ref{fig:sm-gp-ga-yield} are mainly the statistical
uncertainties. The systematic uncertainties arise from the photon
emission probabilities for the decays stemming from the $(\gamma,p)$
(6\%) and $(\gamma,\alpha)$ (8\%) reactions (see
Table~\ref{tb:decayproperties}) and from the full-energy peak
efficiency calculations (2\%).

\section{Conclusions}

Activation experiments of the kind described - especially when
performed with a transport system allowing the study of short
half-lives deliver exact photon strength data. Thus they allow a
verification of data obtained previously by direct absorption
experiments or by detecting the neutrons from the  process. In
Fig.~\ref{fig:sm-gx-parametrization}, the photoneutron cross
sections of the Sm-isotopes as described by the photon strength
parametrization from Junghans $\textit{et al.}$~\cite{Junghans2008}
is shown.

From the figure, it is obvious that the parametrization describes
all the isotopes well except $^{144}$Sm in which the contribution
from $^{144}$Sm$(\gamma,\gamma')$ and $^{144}$Sm$(\gamma,p)$ is of
significant magnitude at low energies close-around S$_{n}$ (see
Fig.~\ref{fig:sm-gn-gp-ga-crosssec}). For all cross section data
from Carlos $\textit{et al.}$, a correction factor of 0.8 based on
Berman $\textit{et al.}$~\cite{Berm87} experiments was applied.
For the $^{144}$Sm cross sections, the additional correction
factor arising from isotopic contaminations was also necessary
(see Sec.~\ref{sect:sm-gnyield}). The yield calculations using
cross sections from the NON-SMOKER code with default nuclear
physics inputs are also shown. It is clear the calculations
incorporating the new parametrization provides a better
description to the experimental data than default NON-SMOKER
results.

The agreement between the parametrization and experimental data is
surprisingly good- taking into account that all parameters (except
$\beta$) were obtained from global fits to all nuclei with $A>70$
(as long as respective data are available for them). The only
non-global parameter, the deformation $\beta$ was directly derived
from the electric quadrupole transition strength to the first
2$^{+}$ state \cite{Raman2001}. The approximation for the
triaxiality $\gamma$ is from the systematics presented by
\cite{Andre93}. The values of $\beta$ and $\gamma$ used for each
isotope are shown in the respective figure. The Lorentzian resonance
integral corresponds to the TRK sum rule and the resonance energies
are calculated using the symmetry energy J=32.7 MeV and surface
stiffness Q=29.2 MeV~\cite{Moll95}.

To summarize, this paper presents the photodisintegration of $
^{144}$Sm to four exit channels. The $ ^{144}$Sm($\gamma$,n)
reaction yield has been compared to the yield calculated using cross
sections from previous photoneutron experiments and a comparison of
the two data sets leads to a conclusion on the inaccuracies in
previous data.  It has been verified that the recommendations based
on Berman $\textit{et al.}$~\cite{Berm87} experiments are necessary
to correct the cross section measurements at Saclay.

The activation yield for all the photodisintegration reactions has
been compared with TALYS and NON-SMOKER models. In general, the
experimental activation yields agree within a factor of 2 to the
simulated yield using statistical model predictions. In view of
the fact, that the codes NON-SMOKER and TALYS are based on the
same theory-the Hauser-Feshbach model, eventual differences
appearing in the model predictions are strongly related to their
nuclear physics inputs. The effect of using different photon
strength functions in the TALYS model calculations has been
discussed here. For the $^{144}$Sm($\gamma$,$\alpha$)reaction, the
statistical model description faces the additional problem of the
possible emission of low energy photons prior to the alpha
emission. In a forthcoming paper the respective data for
$^{144}$Sm will be presented together with results for a different
mass region, and the proper treatment of that process by the two
codes will be examined.

The new proposed phenomenological parametrization describes very
well the photoneutron cross section data for Sm isotopes. A
similar investigation has been performed for the stable even
isotopes of Mo and same effects has been
observed~\cite{Erhard2009}. It is thus worthwhile to point out
again that this new fit only needs a very limited number of fit
parameters to describe the electric dipole strength in all nuclei
with $A>70$, for which nuclear photoeffect data exist.

\section{Acknowledgements}
We thank P. Michel and the ELBE team for providing a stable beam
during activation experiments. We are indebted to J. Klug for the
MCNP simulations of the bremsstrahlung spectra at the
photoactivation site. Special thanks are due to M. K\"ohler and
D. Degering, for their help with the measurements at the
underground laboratory Felsenkeller, Dresden. The technical
assistance of A. Hartmann is gratefully acknowledged.

\end{document}